\documentclass[useAMS,usenatbib,a4paper]{mn2e}
\usepackage[dvips]{graphicx}
\usepackage{graphics}
\usepackage{wrapfig,epsfig,epsf}
\usepackage{amsmath}

\title[Giant pulses of PSR B1937+21]{Statistical and polarization properties of giant
pulses of the millisecond pulsar B1937+21}
\author[V.I.~Zhuravlev et al.]{V.I.~Zhuravlev,$^{1}$\thanks{E-mail: zhur@asc.rssi.ru (VIZ);
popov069@asc.rssi.ru (MVP); vsoglasn@asc.rssi.ru (VAS); kondratiev@astron.nl (VIK);
yyk@asc.rssi.ru (YYK); bartel@yorku.ca (NB); fghigo@nrao.edu (FG)}
M.V.~Popov,$^{1}$\footnotemark[1] V.A.~Soglasnov,$^{1}$\footnotemark[1]
V.I.~Kondrat'ev,$^{2,1}$\footnotemark[1]
\newauthor
 Y.Y.~Kovalev,$^{1,4}$\footnotemark[1]
N.~Bartel$^{3}$\footnotemark[1] and F.~Ghigo$^{4}$\footnotemark[1]\\
$^{1}$Astro Space Center, Lebedev Physical Institute, Profsoyuznaya 
str.84/32, Moscow 117997, Russia\\
$^{2}$Netherlands Institute for Radio Astronomy, P.O. Box 2, 7990 AA Dwingeloo, The
Netherlands\\
$^{3}$York University, Department of Physics and Astronomy, 4700 Keele Street, 
Toronto, Ontario M3J 1P3 Canada\\
$^{4}$National Radio Astronomy Observatory, P.O. Box 2, Green Bank, WV 24944, USA}
\begin{document}

\date{Accepted 2013 January 11. Received 2013 Januaru 7; in original form 2012 October 31}

\pagerange{\pageref{firstpage}--\pageref{lastpage}} \pubyear{XXXX}

\maketitle

\label{firstpage}

\begin{abstract}

We have studied the statistical and polarization properties of giant 
pulses (GPs) emitted by the millisecond pulsar B1937+21, with  
high sensitivity and time resolution. The observations were made in June 
2005 with the 100-m Robert C. Byrd Green Bank Telescope  at S-band (2052-2116 MHz) using 
the Mk5A VLBI recording system, with formal time resolution of 16 ns. The 
total observing time was about 4.5 hours; the rate of detection of GPs was  
about 130 per hour at the average longitudes of the main pulse (MPGPs) and 60 per hour 
at the interpulse (IPGPs). While the average profile shows well-defined polarization behavior,
with  regular evolution of the linear polarization position angle (PA), GPs exhibit random 
properties, 
occasionally having high linear or circular polarization.  Neither MPGPs nor IPGPs
show a preferred PA. The cumulative probability distribution 
(CPD) of GP pulse energy was constructed down to the level where GPs merge with regular 
pulses and noise. For both MPGPs and IPGPs, the CPD follows a power law with a break,
the power index changing from -2.4 at high energy to -1.6 for low energy.
Pulse smearing due to scattering masks the intrinsic shape and duration of the detected GPs.
The smearing time varied during the observing session within a range of a few hundred nanoseconds. 
The measured polarization and 
statistical properties of GPs impose strong constraints on physical models 
of GPs. Some of  these properties support a  model in which GPs are generated by the electric 
discharge caused by magnetic reconnection of field lines connecting the opposite magnetic 
poles of a neutron star.
\end{abstract}


\section{Introduction}

Giant pulses from the millisecond pulsar B1937+21 were first noted by \citet{p1},
and confirmed by \citet{sb}.
\citet{cs} determined the main properties of GPs 
of millisecond pulsars in their study based on 44 min of observation 
with the Arecibo radio telescope at 430 MHz with a time resolution of 1.2 $\mu$s.
They found that GPs were seen in both the main pulse and the interpulse 
components. The GPs were found to be very short in duration. They are delayed by 
40-50 $\mu$s relative to the profile components of the regular emission, and
they were highly circularly polarized, sometimes up to 100\%. \citet{cs}
also established a power law energy distribution with the exponent of $-1.8$ for 
the cumulative energy distribution.
\\ \indent \citet{kt} have published results of multifrequency 
observations of giant radio pulses from B1937+21 observed with the Arecibo radio 
telescope at 430, 1420 and 2380 MHz using a time resolution of 0.38 $\mu$s. They
confirmed very short durations of GPs only restricted by the scattering time
at each frequency. Although the multifrequency observations were not simultaneous,
a mean radio spectrum was estimated to follow a power law with 
exponent $\simeq$ 3.1, somewhat steeper than regular radio emission.

\citet{p2} presented an analysis of simultaneous 
dual-frequency observations of GPs from PSR B1937+21, during 
about 3 hours with the Kalazin 64-m radio telescope at 1420 MHz, and the 
Westerbork Synthesis Radio Telescope (WSRT)
at 2200 MHz. While more than a dozen GPs were detected at Kalazin and 
at WSRT, no events were found to occur simultaneously at both frequencies.
Thus, the instant radio spectra of GPs are subject to deep modulation 
at a frequency scale of about $\Delta\nu / \nu\simeq0.3$.
\citet{b14} presented 
results of observations made at 1650 MHz with the Tidbinbilla 70-m DSS43 
radio telescope with high time resolution. They have detected pulses as strong 
as 65000 Jy with widths $\leq15$ ns, corresponding to a brightness temperature 
of $T_b>5\times10^{39}$ K.

In this paper we present a detailed analysis of 4.5 hours of data obtained 
in June 2005 with the GBT radio telescope in a 2052-2116 MHz frequency
band using dual circular polarization.

In Section~\ref{obs} we describe the observations and data reduction. In Section~\ref{polar}
we discuss our approach to calibration of polarization and we 
present the polarized profile of our calibrator PSR B1929+10.
The average profile of B1937+21 in full polarization is described in Section~\ref{prof}. 
In Section~\ref{detect} we explain the technique of GPs detection, and in the following sections
we present our results on polarization (Section~\ref{GPs}), statistics
(Section~\ref{stat}), and scattering (Section~\ref{scpr}). Our conclusions are summarized in 
Section~\ref{disc}.

\section[]{Observations}
\label{obs}
The observations were conducted in June 2005 with the 100-m Robert C. Byrd Green Bank  
Telescope  (GBT) at S-band (2052-2116 MHz) using a VLBI Mk5A  terminal. 
Two polarization channels  (RCP and LCP) were recorded 
with the following frequency setup: 2068.0 and 2100.0 MHz sky frequencies, each
 with USB and LSB 16-MHz subbands sampled at  the Nyquist 
frequency (31.25 ns sampling time) with 2-bit digitizing. 
Thus, the configuration provided four 
16-MHz conjugate bands per polarization potentially enabling 16 ns time resolution in each
polarization channel.
However we did not use such ultimate time resolution 
since scatter broadening was found to be about 100 ns, completely
masking the intrinsic time shape of giant pulses (see Section~\ref{scpr} for details).
 The calibration was performed by injecting a
noise diode signal with level 2.1 K, and by observations of bright 
continuum sources 3C286 and 3C399.1. 
Conversion to 
flux density (Jy) used  noise diode amplitudes calibrated on 3C286 and
3C399.1 with flux densities from \citet{ba}.
The System Equivalent Flux Density (SEFD) was estimated to be 
 11.2 Jy, and the value was used to convert our measurements to Jy throughout
the paper. We believe that our flux calibration is within 10\% in relative accuracy. 
Pulsar B1929+10 was observed several 
times at different hour angles for polarization calibration. Data, amounting to
about 2TB total, were recorded 
on Mk5 diskpacks, copied to regular hard disks, and transferred to the 
processing site.

Data reduction started
with Mk5a data decoding with the ``mk5decode'' routine composed by V.Kondratiev 
\citep{b1}.  The routine selects channels and converts 2-bit 
codes into floating point numbers  -3, -1, +1, +3. This simple decoding was accompanied 
with corrections for two-bit statistics by the rules originally proposed 
by \citet{b2}.  Such a correction is especially important in 
pulsar observations when a strong pulse is expected. For a strong pulse the threshold 
for discrimination between 1 and 3 does not correspond to a nominal 
1 sigma level. 

Efforts were made to remove several strong interfering signals in the radio spectra by 
substituting pseudo-random noise in the infected regions.  The bandpass shape for 
each subband was individually corrected to provide a more efficient application 
of the predetection dispersion removal technique originally proposed by \citet{b3}. 
Some details of practical usage of the technique can be found in our previous 
publications: \citet{b4, b5, b6}. Dispersion removal 
was carried out separately for each frequency and polarization channel (with 
31.25 ns sampling time), but additional time shifts were simultaneously 
applied to different frequency channels, to align to the frequency 
of 2100 MHz.  This last measure allows averaging data synchronously in all 
frequency channels, providing better sensitivity. The averaging became only 
possible after reduction of the raw data to real Stokes parameters.

\section{Polarimetric calibration using PSR B1929+10}
\label{polar}
Raw observed Stokes parameters were calculated 
after dedispersion, using the following expressions \citep{b4}:
\begin{eqnarray}
I_m & = & W_{\rmn{r}}W_{\rmn{r}}+W_{\rmn{i}}W_{\rmn{i}}+X_{\rmn{r}}X_{\rmn{r}}+X_{\rmn{i}}X_{\rmn{i}}\nonumber\\
V_m & = & W_{\rmn{r}}W_{\rmn{r}}+W_{\rmn{i}}W_{\rmn{i}}-X_{\rmn{r}}X_{\rmn{r}}-X_{\rmn{i}}X_{\rmn{i}}\\
Q_m & = & 2(W_{\rmn{r}}X_{\rmn{r}}+W_{\rmn{i}}X_{\rmn{i}})\nonumber\\
U_m & = & 2(X_{\rmn{r}}W_{\rmn{i}}-X_{\rmn{i}}W_{\rmn{r}})\nonumber
\end{eqnarray}
where $W_{\rmn{r}}$, $W_{\rmn{i}}$, $X_{\rmn{r}}$ and $X_{\rmn{i}}$ represent the real and 
imaginary components of the complex 
analytic signal in the RCP(W) and the LCP(X) polarization channels. 

\begin{figure}
\includegraphics[width=84mm,trim=0.0cm 0cm 0cm 1.cm,clip]{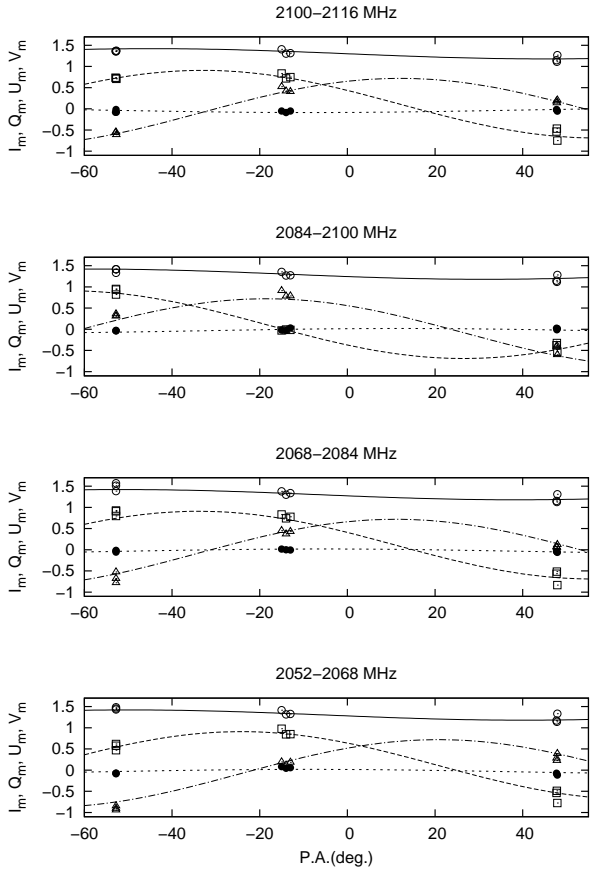}
 \caption{Measured Stokes parameters from PSR B1929+10 in four 16 MHz subbands, normalized by the invariant 
 interval  $\sqrt{I^2_m-Q^2_m-U^2_m-V^2_m}$ and plotted as a function of parallactic angles.
 Each data point represents the average of about 300 pulsar periods.  Here, $I_m$ is denoted 
 by open circles, $Q_m$ by squares, $U_m$  by triangles, and $V_m$ by closed circles. 
 Stokes parameters were estimated by fitting sine functions with a nonlinear least squares method. 
 Solid, dashed, dash-dotted, dotted lines represent the approximations for 
 $I_m$, $Q_m$, $U_m$, $V_m$ correspondingly.}
 \label{fig.1}
\end{figure}

Observations of the pulsar PSR B1929+10 were used to estimate the instrumental 
cross-coupling terms of the GBT. 
The observed 
Stokes parameters at maximum intensity in the average profile 
of PSR B1929+10 are displayed in  Figure~\ref{fig.1},  plotted versus parallactic 
angle for each frequency band. 

\begin{figure}
\includegraphics[width=0.48\textwidth,angle=0,trim=0.0cm 0.cm 0cm 0.cm,clip]{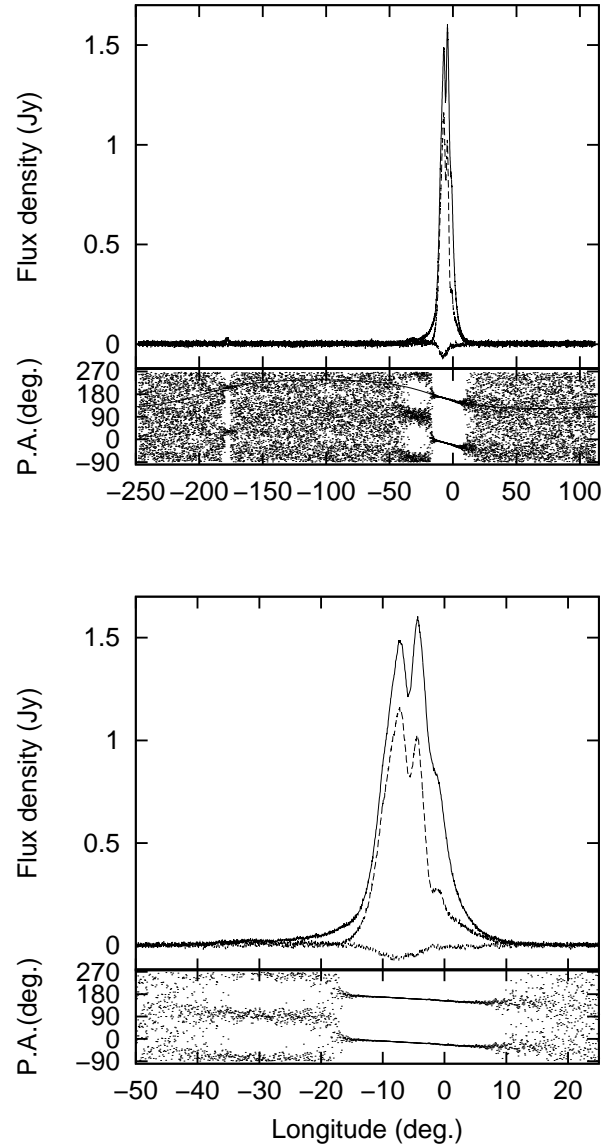}
 \caption{Pulse profiles for PSR B1929+10 after polarization calibration. 
 The solid  line  represents the total intensity {\it I}, the shaded line is the 
 linear intensity $L=\sqrt{Q^2+U^2}$,  the shade-dotted line is the circular intensity 
 {\it V}.  The position angle $P.A.=0.5\arctan(U/Q)$ is plotted 
 twice for clarity.  The upper portion of the figure shows the pulse profiles for the full period 
 of the pulsar, and the bottom one shows only the main pulse. } 
 \label{fig.2}
 
\end{figure}
Following the recommendations of 
\citet{b8} we use the following  relation
\begin{equation}
S_{\rmn{obs}}(I_m,Q_m,U_m,V_m)=\mid M_{\rmn{i,j}}\mid S_{\rmn{true}}(I,Q,U,V)    
\end{equation}
It was possible to decompose the original matrix  
$\mid M_{\rmn{i,j}}\mid $ as a combination of 
two matrices: a time-independent matrix $\mid T_{\rmn{i,k}}\mid $,  
and a matrix $\mid B_{\rmn{k,j}}\mid $ describing 
the rotation of the telescope while tracking the radio source
\begin{equation}
\mid B_{\rmn{k,j}}\mid = 
\left| 
\begin{array}{crrc}
1 & 0 & 0 & 0 \\
0 & \cos 2\beta & -\sin 2\beta & 0 \\
0 & \sin 2\beta & \cos 2\beta & 0 \\
0 & 0 & 0 & 1 \\
\end{array}
\right| 
\end{equation}
where $\beta$ is the parallactic angle. 
In the general case one needs to find all 16 unknown coefficients of 
$\mid T_{\rmn{i,k}}\mid$,  
but \citet{b7} simplified the procedure and reduced the number of unknown 
coefficients to 5. We applied the combined method of Turlo and McKinnon using our 
calibration measurements on PSR B1929+10 presented in  Figure~\ref{fig.1}.  
We estimate depolarization $D_p=\frac{\sqrt{T^2_{12}+T^2_{13}}}{\mid T_{11}\mid}$,
representing the linear polarization which appears as total intensity. 
The instrumental polarization $A=\frac{\sqrt{T^2_{21}+T^2_{31}}}{\mid T_{11}\mid}$
represents the total intensity which
appears as linear polarization.
Values of $D_p$ and $A$ have 
a value of 11\%. These  values were found to be the same in all four 
frequency channels while the instrumental phases are different.
A similar cross-coupling, about 6.8\%,  occurs between the circular and linear polarization.
We do not take into account the rotation measure correction between frequency bands which 
constitutes only $2^{\circ}$.
The recovered 
polarization profile of the PSR B1929+10 is shown in  Figure~\ref{fig.2}. The profile was
 formed by averaging about 2800 pulses at parallactic angles of $-54^\circ$, $-17^\circ$, 
 and $48^\circ$.     In the frame of the rotating vector model (RVM) \citep{b10}  
 one can estimate geometric parameters of the neutron 
star rotation by using the PA variation curve for the main pulse and interpulse 
longitudes (not shown in Figure~\ref{fig.2}). 
The curves indicate the fits to  RVM:
 $P.A.(\phi)=\arctan\left[\frac{\sin(\alpha)\sin(\phi-\phi_{\circ})}{\sin(\xi)\cos(\alpha)-\cos(\xi)\sin(\alpha)\cos(\phi-\phi_{\circ})}\right]+const, $ 
 where $\alpha$  is an 
angle between the spin axis and the magnetic axis,  $\beta$ is an angle between the 
magnetic axis and the line of sight from Earth, and $\xi=\alpha+\beta$ is the angle between the rotation 
and the the line of sight. 
Our estimates are $\alpha=54\degr$ and $\beta=43\degr$. 
The profile is 
in  good agreement with the published data (see, for example, 
\citet{b9}).

According to the theory of radiation transfer developed
by \citet{bf}, the mean pulse for this pulsar
is formed by the extraordinary mode (the signs of the PA
derivative and circular polarization are the same).

\section{Average profile of PSR B1937+21}
\label{prof}
\begin{figure}
\includegraphics[width=0.47\textwidth,angle=0,trim=1.4cm 0.cm 0cm 0.cm,clip]{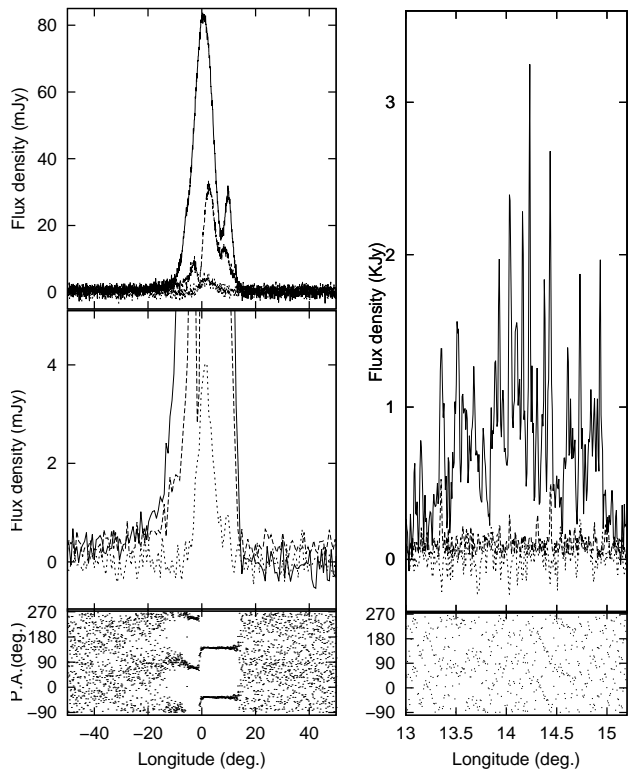}
 \caption{Regular emission (left) and averaged GPs profile  (right) for the main pulse of the
 PSR B1937+21  after time smoothing
 over 312.5 nc (10 points). Full intensity -- solid line, linear intensity -- shaded line, and
 circular intensity -- dotted line. 
 } 
 \label{fig.3}
\end{figure}

\begin{figure}
\includegraphics[width=0.49\textwidth,angle=0,trim=1.0cm 0.cm 0cm 0.cm,clip]{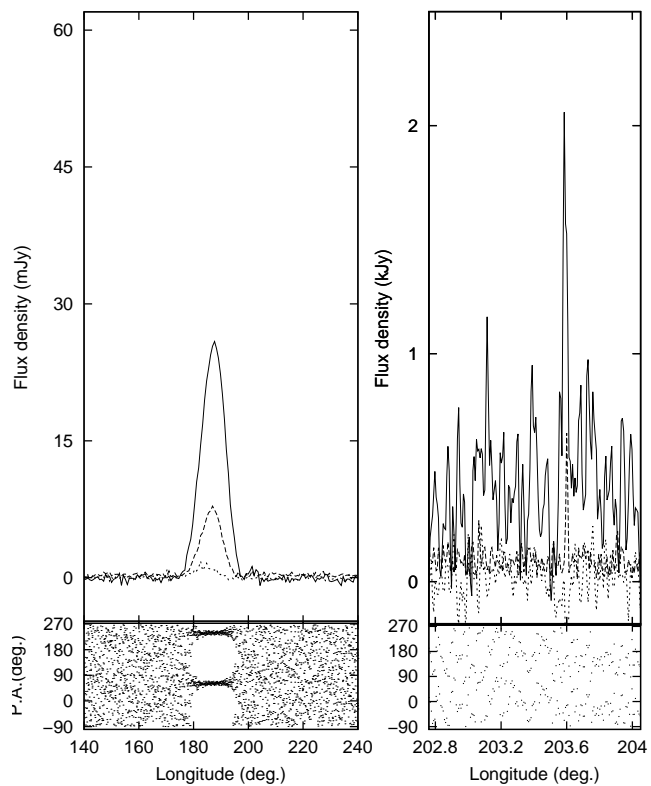}
 \caption{Regular emission (left) and averaged GPs profile  (right) for the interpulse of
 PSR B1937+21 after time smoothing 
 over 312.5 nc (10 points).  All other details are the same as for figure~\ref{fig.3}. 
 } 
 \label{fig.4}
\end{figure}

Polarization properties of the average profile for the millisecond pulsar B1937+21 
were presented and discussed in several publications (\citet{b11, ts, stc, kp, b13}). 
In our study we have the best time resolution and good sensitivity compared 
with the above mentioned investigations. The average polarization profile obtained in 
our study with the GBT at S-band is shown in Figures~\ref{fig.3} and \ref{fig.4} with time resolution 
of 312.5 ns (10-points time smoothing), constructed by summing all four 
frequency bands. The ephemeris and value of the dispersion measure 
DM=71.0398~pc~cm$^{-3}$   
were taken from \citet{MHTH} and corrected by the observed time of arrival of GPs
(DM=71.029~pc~cm$^{-3}$).


In general our results agree with previous studies, but we would like to note 
some peculiarities. We confirm the nearly flat shape of PA with longitude 
over the major portion of the main pulse and over the full range of longitudes at the 
interpulse. Note, that the nearly 90-degree jump in the PA curve coincides with the sudden 
increase of circular polarization, and the flat portion of the PA curve covers 
continuously the tail of the whole main component including the second distinct component of
the main pulse. In fact, the observed jump in the PA is not equal to $90\deg$, but is 
definitely closer to $80\degr$; this peculiarity was also found by  
\citet{b13}. We did distinguish very well the presence of an extended weak 
component preceding the main one. 

It is interesting to stress that the PA curve for this pulsar is flat.
According to \citet{bf}, this is typical of
millisecond pulsars, in which the polarization characteristics are
formed near the light cylinder, where the magnetic field of a
neutron star is close to uniform.

\section{Detection of giant pulses}
\label{detect}
Searches for giant pulses were conducted in full intensity I averaged over all 
frequency bands (8 channels altogether), with the resulting root mean square 
deviation (RMS) of $\sigma=SEFD/\sqrt(8)=4$Jy. Such a signal follows $\chi^2$ statistics 
with 16 degrees of freedom.  It is well known that GPs from the millisecond pulsar B1937+21 
are very short  \citep{b14}, but still they  may have slightly 
different durations, and  we tried  several cases of time averaging by 1, 3, 5, 
and 7 points to get the best SNR for a given GP. 
For each case we applied a particular threshold for the 
detection. The values of the thresholds used are given in Table~\ref{tab1}.
\begin{table}
\caption{GP detection thresholds for various signal times}
\label{tab1}
\begin{tabular}{@{}lrrrr}
\hline
Averaging       & 1     & 3    & 5      & 7     \\
\hline
Duration (ns)        & 31.25 & 62.5 & 151.25 & 218.75 \\
GP threshold in $\sigma$  & 13    & 11   &  10    &  9      \\
GP threshold in Jy  & 52    & 25   &  18    &  13      \\
\hline
\end{tabular}

\medskip
The thresholds were calculated for the expected $\chi_P^2(\nu)$ distribution
$P(\sigma_i>\sigma_{th})=\exp(-\frac{\sigma_{th}}{2})
\sum\limits_{j=0}\limits^{j=\nu/2-1}\frac{1}{j!}\left(\frac{\sigma_{th}}{2}\right)^j$ 
by the condition of $P(\sigma_i>\sigma_{th})<3\cdot10^{-13}$ resulting in less than 1\%
false detections in strong GPs with energies greater than 2 Jy$\cdot \mu$s (see
Section~\ref{stat}). 
\end{table}
 In total there were detected 
597 and 282 GPs at the longitudes close to the main pulse (MPGP)
and the interpulse (IPGP) correspondingly. Time intervals between the middle 
of the GP window and the longitude of the maximum of the corresponding component of
the average profile were found to be 59.2 and 67.3 $\mu$s,  in  very good agreement 
with the values given by \citet{b14} at 1650 MHz (58.3 and 65.2 $\mu$s).
Therefore, there is no notable expansion in the separation of regular and GP profiles 
between frequencies of 1650 and 2100 MHz.

\section{Polarization properties of giant pulses.}
\label{GPs}
\begin{figure}
\includegraphics[width=0.60\textwidth,angle=0,trim=0.0cm 0.0cm 0.0cm 2.8cm,clip]{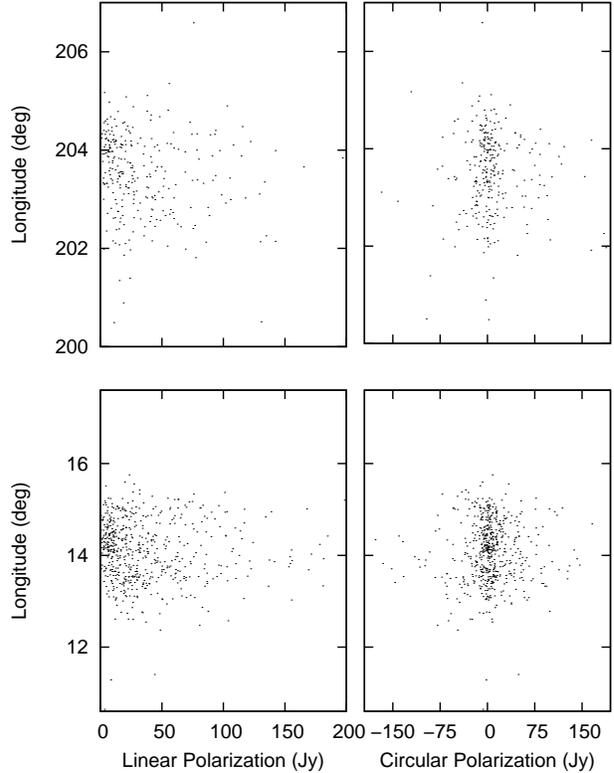}
 \caption{ Distributions of circular intensity (right) and linear polarized intensity (left) 
 RPs for  PSR B1937+21. Versus the longitude of pulses for main pulses (down) and interpulse 
 (up).} 
 \label{fig.5}
\end{figure}

Using the technique described in Section~\ref{polar} we calculated 
corrected Stokes parameters for every frequency band. 
Additionally, we averaged Stokes parameters over five 
points near every GPs maximum. Thus every value is an average over 20 points
(four in frequency and five in time). In Figures~\ref{fig.3} and \ref{fig.4}
 we compare average 
profiles in full polarization for regular and GPs emission. While the regular 
emission shows well defined behaviors of Stokes parameters with the
longitude of the average profile, profiles of GPs do not show any regular polarization.
Only  strong GPs (with total intensity 
$I_{MPGP}>21\sigma (84 Jy)$ and $I_{IPGP}>19\sigma (76 Jy)$) 
were used for this study.
Individual GPs have high linear or circular polarization 
of both signs. However, there  is no dependence of Stokes parameter on 
longitude, or on the intensity of the GPs. We have analyzed the distributions 
of GPs Stokes parameters, and have found them to be random.
An example of longitude distribution is given in Figure~\ref{fig.5}.

\section{Energy  distribution}
\label{stat}
\begin{figure}
\includegraphics[width=0.50\textwidth,angle=0,trim=0.0cm 0.cm 0cm 0.cm,clip]{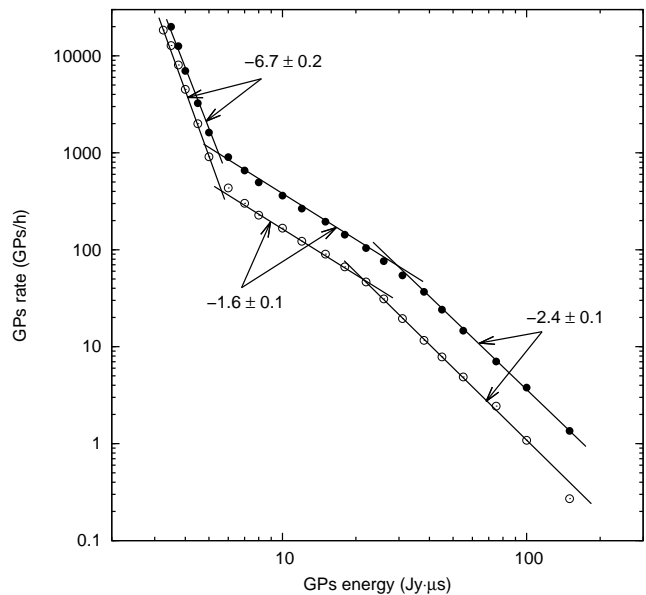}
 \caption{Cumulative probability distribution of the energies of GPs in the main 
 pulse (filled circles) and interpulse windows (open circles). Solid lines represent 
 least-squares linear fits.} 
 \label{fig.6}
\end{figure}

The energy of each GP was calculated as 
\mbox{$E=\underset{N=1,3,5,7}{\max}{(\delta t\sum\limits_{i=1}F_i)}$} with F being flux density Jy, 
and $\delta t$ is the sampling interval (31.25 ns). We integrated over 
windows of 1, 3, 5 or 7 points where the GP showed the greatest SNR.  
To compensate for false detection at low energies, we conducted 
an identical calculation in the off-pulse window. Finally we subtracted the normalized 
number of detections in the off-pulse window from the number detected in the
on-pulse window for each energy interval (for weak pulses with energy below 
6 Jy$\cdot\mu$s
such corrections were essential). After these corrections the 
cumulative probability distribution (CPD) was obtained separately for the
main pulse and the the interpulse GPs. The CPDs are shown in Figure~\ref{fig.6}. 
The CPDs were approximated by power law functions with  different exponents in 
different energy ranges, the values were found to be similar for MPGPs and IPGPs. 
The exponent for the high
energy GPs is $-2.4\pm0.1$, while the exponent for low energy GPs is $-1.6\pm0.1$. 
Values of the energy $E_{break}$ are 30.2 and 20.1 Jy$\cdot\mu$s for the MPGPs 
and IPGPs respectively.
At  very 
low energies we detected pulses of regular emission at the trailing edge of the
average profile with the Gaussian distribution,  where the CPDs have a fast increase.

Let us consider that the difference in $E_{break}$ is caused by the beaming effects, 
i.e. we have $2/3$ attenuation in IPGPs. Then we shall have the number of detections
for a given flux density value $N_{IPGP}=\left(\frac{2}{3}\right)^{-5/2}N_{MPGP}$
for the strong region of CPD. This is exactly the case observed.

A similar break in CPD was found by \citet{p22} for the GPs from the Crab
pulsar in their analysis of 3.5 hours of observation with the WSRT at 1200 MHz.
They gave a similar explanation (beam attenuation) for the CPD peculiarities for IPGPs and MPGPs.
 
\section{Scattering parameters}
\label{scpr} 
\begin{figure}
\includegraphics[width=0.50\textwidth,angle=0,trim=0.0cm 0.cm 0cm 0.cm,clip]{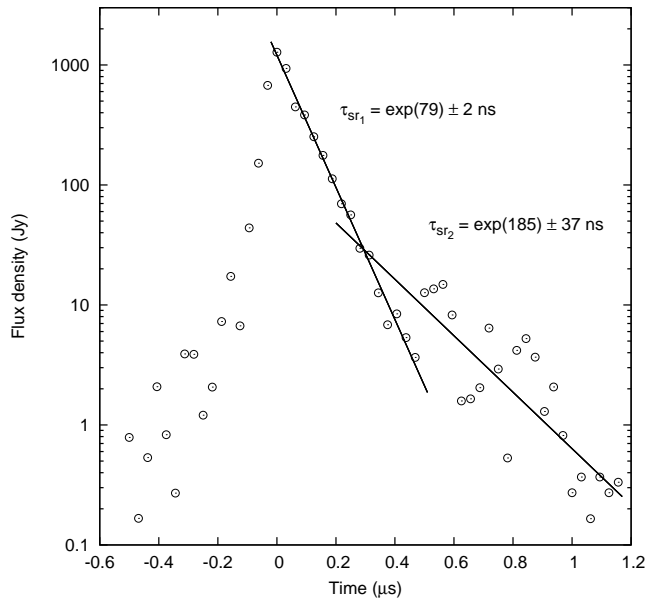}
 \caption{The profile of the strongest  GP of B1937+21  (open circles). Solid lines show 
 the least-squares fits to the exponential scattering tail of the profile,  giving 
 estimated scattering times of $\tau_{sc_1}=79\pm2$ and $\tau_{sc_2}=185\pm37$  ns. 
 } 
 \label{fig.7}
\end{figure}

\begin{figure}
\includegraphics[width=0.51\textwidth,angle=0,trim=.0cm 1.cm 0.cm 2.cm,clip]{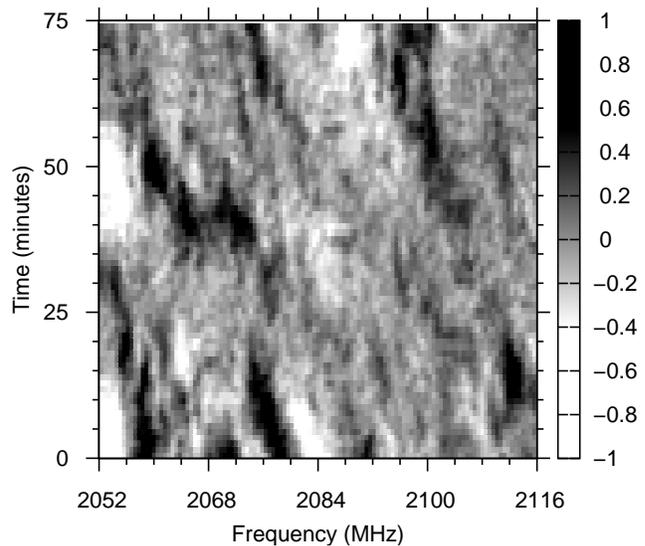}
 \caption{Dynamic spectra of selected observation data with high GP activity.
 } 
 \label{fig.8}
\end{figure}

The scattering of radio waves is manifested in the following effects: pulse 
broadening, intensity variations, distortion of radio spectrum, angular 
broadening. To estimate pulse broadening we selected several strong GPs 
not showing any evident intrinsic structure, and have measured the value of 
$\tau_{sc}$ in the exponential tail of the pulse shape $Y(t)=Ae^{-t/\tau_{sc}}$.
Figure~\ref{fig.7} illustrates the technique.  Values of
$\tau_{sc}$ ware measured to be $79\pm2$ and $185\pm37$ ns.   
While the short time-scale is based on the pulse portion with good signal-to-noise ratio
($SNR > 10$), the long time scale was estimated in the portion of the pulse merging with noise
($\sigma=4$Jy), but we consider it to be real since the approximation was done over 
about 1 $\mu$s time interval, thus providing averaging with the resulting $\sigma$
near 1 Jy.
The pulse broadening time $\tau_{sc}$ must be correlated with the decorrelation bandwidth 
$\Delta\nu_d$ in the radio spectrum through the relation $2\pi\tau_{sc}\Delta\nu_d=C$ with
$C$ being a constant close to 1.0. With this relation we can expect to find two frequency
scales for $\Delta \nu_d$ of about 2 MHz and 0.85 MHz in the radio spectrum.
 To measure the  $\Delta\nu_d$ we constructed the dynamic
spectrum over the total 64 MHz band (B). The border regions between conjugate 16 MHz
sub-bands were filled with random noise with corresponding mean value and 
variance. The example of such a dynamic spectrum is presented in Figure~\ref{fig.8}.
To obtain values for scintillation bandwidth and scintillation time we use 
a 2-dimensional autocorrelation function (ACF) of the dynamic spectrum. Frequency 
and time sections were approximated by a combination of Gaussians. It
was found that there are three frequency scales in the diffraction pattern with 
$\Delta\nu_{d_1}=2220\pm 20$, $\Delta\nu_{d_2}=690\pm 15$ and $\Delta\nu_{d_3}=110\pm 5$ kHz
at a half width level. For the region of low activity the scintillation time $\Delta t_{sc}$ was 
fitted with only one Gaussian ($\delta t_{sc}=6$~min), and two components 
($\delta t_{sc}$ = 7 and 2 min) were found for the region
of high activity. The results are presented in  Table~\ref{tab2}.

\begin{table}
\caption{Amplitudes and widths for time and frequency sections of the ACF of dynamical spectra
for two stages of activity GPs. (Only statistical fitting errors are given.
)}
\label{tab2}
\begin{tabular}{@{}lrr} 
\hline
& \multicolumn{2}{c} {Time scales} \\ 
                & Region of high activity         & Region of low activity  \\
\hline
$A_1\cdot 10^{-3}$  & $63.4\pm 0.2$    & $103.3\pm 0.6$ \\ 
$A_2\cdot 10^{-3}$  & $14.7\pm 0.2$    &                 \\ \hline
$\Delta t_{sc_1}$ (min)     & $7.1\pm 0.8$    & $6.1\pm 0.5$  \\
$\Delta t_{sc_2}$          & $2.1\pm 0.4$     &               \\ \hline
&  \multicolumn{2}{c} {Frequency scales}\\ 
                & Region of  high activity         & Region of  low activity       \\
\hline
$A_1\cdot 10^{-3}$   & $57.0\pm 0.5$  & $52.1\pm 1.5$ \\ 
$A_2\cdot 10^{-3}$   & $15.3\pm 0.6$  & $39.0\pm 1.5$  \\
$A_3\cdot 10^{-3}$   & $6.2\pm 0.8$   & $12.5\pm 0.5$  \\ \hline
$\Delta\nu_{d_1}$ (kHz)    & $2220\pm 20$  & $3650\pm 60$  \\
$\Delta\nu_{d_2}$    & $690\pm 15$    & $1740\pm 30$ \\ 
$\Delta\nu_{d_3}$    & $110\pm 5$      & $500\pm 10$ \\ \hline
\end{tabular}
\end{table}

If we disregard the multiple frequency structure, we can use the
value of $\Delta\nu_d=1.7$ MHz. With this value of $\Delta\nu_d$ we have a
scintillation intensity modulation in the total $B=64$ Mhz band being significantly
reduced ($\Delta\nu_d/B=0.025$). Therefore our analysis of energy distribution was not  
affected by the scintillations.

We noticed two time intervals of GP activity that may be connected with scintillation effects.
The region of high activity was about one hour duration, and
the region of low activity was about three hours.

\section{Discussion and Conclusions}
\label{disc}
Let us review our findings:

1) The observed durations of GPs from PSR B1937+21 reflect mainly scatter
broadening of about 100 -- 200 ns, while the intrinsic time width was unresolved 
($\la$ 30 ns).

2) Our analysis of polarization properties of GPs revealed random 
behavior of polarization with occasional pulses showing high linear 
polarization as  well as high circular polarization of both signs, in
contrast to the average profile of regular radio emission. 
Such polarization properties impose strong constraints for theoretical models 
explaining physical nature of GPs.

3) The cumulative probability distribution in energy was obtained based on a large volume of 
statistical data down to 3 Jy$\cdot \mu$s. 
The CPD demonstrated that there is no cutoff at low energies, 
i.e. the weakest GPs merge smoothly with the regular component of radio emission.

4) The CPD has a definite break in the power law exponent at energies equal
to 30.2 and 20.1 Jy$\dot\mu$s for the MPGPs and IPGPs respectively. At these 
energies the exponent changes its value from $-1.6 \pm 0.1$ to $-2.4\pm 0.1$, i.e.
approximately by $-1$.
The same behavior was found for the CPD Crab pulsar GPs, as published by \citet{p22}.
Thus, an observed break in the exponent of the power law dependence of the CPD seems to be 
intrinsic to the mechanism of the generation of giant radio pulses.

The above mentioned properties fit very well the explanation given by \citet{is}
in his model of origin of GPs as a result of the reconnection of the last open/close 
magnetic field line (MFL). In a case of nearly perpendicular rotation (magnetic axis is 
perpendicular to the axis of rotation) the last open MFL, when reconnected, will 
join regions of the polar caps with different signs of electric potential $\psi$
thus leading to a strong electric discharge and plasma particle creation and
acceleration. Istomin proposes specific maser amplification of plasma waves
induced by a two-stream instability. 
The model expects large circular components in polarization of GPs. What is interesting,
Istomin's model predicts a power law dependence for the CPD with the exponent  $-3/2$ or
$-5/2$ depending on the power density of plasma generated inside the discharge tube.
Thus, such a transition from one energy state to another may be 
accompanied by a break point in CPD.

\section*{Acknowledgments}
This work was based on GBT observations conducted within the project GBT05B-038.
Our study was supported by the Russian Foundation for Basic Research (project code
10-02-0076) and the Basic Research Program of the Presidium of the Russian
Academy of Sciences on "The Origin, Structure, and Evolution of Objects in the
Universe". The National Radio Astronomy Observatory is a facility of the National 
Science Foundation operated under cooperative agreement by Associated Universities, Inc.
At the time of observations Y.Y. Kovalev was an NRAO Karl Jansky  fellow. 
We thank V.~Beskin and Ya.~Istomin for helpful discussion.

\appendix

\bsp

\label{lastpage}

\end{document}